\documentstyle[12pt]{article}
\begin{document}

\title {{\bf Measuring the strength of dissipative inflation}}

\author{Luis Herrera\thanks{ Also at UCV, Caracas, Venezuela; E-mail address:
lherrera@gugu.usal.es}
 , Alicia Di Prisco\thanks{On leave from Universidad Central de Venezuela,
Caracas, Venezuela}\\
Area de F\'{\i}sica Te\'orica. Facultad de Ciencias.\\ Universidad de
Salamanca. 37008 Salamanca, Spain.\\
and\\
Diego Pav\'{o}n\thanks{E-mail address: diego@ulises.uab.es}\\
Departamento de F\'{\i}sica, Facultad de Ciencias,\\
Edificio Cc, Universidad Aut\'{o}noma de Barcelona,\\
08193 Bellaterra, Spain.
}

\date{}
\maketitle

\begin{abstract}
We comment on recently proposed dissipative inflationary models.
It is shown that the strength of the inflationary expansion is related to
a specific combination of thermodynamic variables which is known to
measure the instability of self--gravitating dissipative systems.
\end{abstract}

\newpage
In recent years different authors have considered the possibility
that inflation could be driven by dissipative fluid effects
(see \cite{PaBaJo}--\cite{MaGoMa} and references therein),
or somehow equivalently by particle production as a result of the
interaction between the quantum vacuum and the gravitational field.
Recently the comparison between the cosmological consequences of both
processes has been studied in detail \cite{zimdahl}. A phase of
accelerated expansion shortly after the big-bang has been invoked
many times as a mechanism able to get rid of the horizon and flatness
problems that beset the standard cosmological model \cite{kolbturner}.

It is the purpose of this note to relate the ``strength'' of the
inflationary phase of those models to a given parameter formed by
a specific combination of hydrodynamical variables. This parameter
has been shown to affect critically the evolution of
self--gravitating dissipative objects
\cite{HeMa98}--\cite{HeMa97}.

The study of the departure of dissipative systems from hydrostatic
equilibrium has shown that the aforesaid parameter $\alpha$ enters
the equation of motion of any fluid element in such a way that the
inertial mass density term is multiplied by a factor that vanishes
for a certain value of that parameter indicating therefore
the existence of a critical point.
In general, the ``effective'' inertial mass density decreases with the
increasing of $\alpha$. In some cases (pure shear or bulk viscosity
\cite{HeMa98}) the critical point is well beyond the border where the
causality requirements are violated. In others (pure thermal conduction
\cite{Heetal97,HeMa97}) the latter requirements are violated slightly
below the critical point. However, in the general case (heat conduction
plus viscosity)
\begin{equation}
\alpha = \frac{1}{(\rho + p)} \left(\frac{\zeta}{2 \tau_{\zeta}} +
\frac{\kappa T}{\tau_{\kappa}} + \frac{2 \eta}{3 \tau_{\eta}}\right)
\, ,
\label{alfag}
\end{equation}
it appears that causality may break down beyond the
critical point \cite{HeMa98}. In the above expression $\rho$ and $p$
are the energy density and equilibrium pressure of the fluid, respectively;
$\zeta$, $\kappa$ and $\eta$ denote the transport coefficients of bulk
viscosity, heat conduction and shear viscosity, respectively; and
the three different $\tau$ stand for the corresponding relaxation times.
Finally $T$ indicates the fluid temperature. Here we shall not consider
anisotropic cosmological models and so the shear coefficient will not
appear.

Our motivation is two fold: on one hand we wish to delve deeper into
the physical meaning of $\alpha$ and, on the other hand, we
would like to provide a ``control'' parameter for the strength of
expansion in any given inflationary model driven by dissipative
processes.

We begin by considering a model, recently proposed by Maartens and
collaborators \cite{MaGoMa} of
an inflationary solution with causal heat flux.
They start from the inhomogenous shear--free model of Modak which in
comoving coordinates has the line element \cite{Modak},
\cite{Krasinski}
\begin{equation}
ds^2 = - \left[1 + M(t) r^2\right]^2 dt^2 + a^2(t) \left[dr^2 + r^2
\left(d\theta^2 + \sin^2{\theta} d\phi^2\right)\right].
\label{le}
\end{equation}
\noindent
Next, they assume that $M=M_0$ and $H\equiv\dot{a}/a=H_{0}$, are both
positive constants and that the stress--energy tensor  of the fluid
can be written as
\begin{equation}
T_{ab} = (\rho + p) u_{a} u_{b} + p g_{ab} + 2 q_{(a} u_{b)} \, ,
\label{Tab}
\end{equation}
\noindent
where $q_{a} =(q/a) \delta^{1}{\,}_{a}$, $q$ being a covariant scalar
measure of the energy--flux magnitude.
Then the expressions for the matter variables which
follow from field equations are
\begin{equation}
\rho=\frac{3H_0^2}{\left(1+M_0 r^2\right)^2},
\label{ro}
\end{equation}
\begin{equation}
p=\left[\frac{4M_0}{a_0^2 \left(1+M_0 r^2\right)}\right]e^{-2H_0 t} -
\rho \, ,
\label{p}
\end{equation}
\begin{equation}
q=- \left[\frac{4M_0H_0 r}{a_0 \left(1+M_0 r^2\right)^2}\right]
e^{-H_0 t}\, ,
\label{q}
\end{equation}
\noindent
where the fact that $a = a_{0} e^{H_{0}t}$ was used.
Also, in oder to satisfy the Israel--Stewart causal transport equation
\cite{IsSt}, \cite{ddj} they
assume
\begin{equation}
\tau_{\kappa}=\left(1+M_0 r^2\right) H_0^{-1}
\label{tau}
\end{equation}
\begin{equation}
T=\frac{U(t)}{1+M_0 r^2} \,  \qquad (U(t) > 0).
\label{T}
\end{equation}

In the model of Maartens {\em et al.} $T$ decreases radially outward
notwithstanding the heat--flux is directed inward. This is so
because the thermal effect concomitant to the fluid acceleration
$\dot{u}_{\alpha} = 2 M_{0} r (1+M_{0}r^{2})^{-1} \delta^{1}{\,}_{a}$
dominates over the temperature gradient \cite{MTW}.

By virtue of (\ref{ro}), (\ref{p}), (\ref{tau}) and (\ref{T}), equation
(\ref{alfag}) becomes into
\begin{equation}
\alpha=\left(\frac{\kappa U(t)}{12}\right) \left(\frac{\Theta}{M_0}\right)
a_0^2 e^{2 H_0 t} ,
\label{al}
\end{equation}
where the fluid expansion rate is given by
\begin{equation}
\Theta \equiv u^{a}{\,}_{;a} = \frac{3 H_0}{1+M_0 r^2} .
\label{Te}
\end{equation}

Now, it is clear that for larger values of $a_0$ and $H_0$, and smaller
values of $M_0$
(which implies larger values of $\Theta$), the inflationary expansion will
be stronger.
On the other hand, as immediately follows from (\ref{al}), it also implies
larger values of $\alpha$.
Therefore, stronger inflationary expansions are related to larger values
of $\alpha$ (for fixed $U(t)$). This result reinforces the physical meaning
of $\alpha$ as a measure of the instability of the system
(see \cite{HeMa98}--\cite{HeMa97}, \cite{HeMaA}, \cite{HeDP}). All this
suggests that $\alpha$ can be used as a ``control parameter''
of the strength of expansion in a given inflationary model driven by causal
heat flux.

As we shall see the previous comment about the potential role of $\alpha$
may be extended to inflationary models driven by viscous stresses with no
heat fluxes.
Indeed, in the case of inflation driven by bulk viscosity the generated
entropy may be written as (see equation (38) in \cite{MaMe}
\footnote{Our $\alpha$ should not be counfused with theirs.})
\begin{equation}
S\approx\frac{4\alpha-1}{2\alpha} \, ,
\label{en}
\end{equation}
where $\alpha$ is now given by (\ref{alfag}) specialized
to the $\kappa = \eta = 0$ case \cite{HeMa98}.
Equation (\ref{en}) shows that the generation of entropy
increases with $\alpha$.
Also observe that in the family of models presented in \cite{Ma},
it appears that
\begin{equation}
\alpha=\frac{1}{2\gamma}
\label{alg}
\end{equation}
where as usual $\gamma$ is the adiabatic index and enters the
equation of state of the fluid
\begin{equation}
p=(\gamma-1) \rho .
\label{g}
\end{equation}
Thus for the case of ultrarelativistic particles ($\gamma=4/3$) we have
$ \alpha= 3/8 $, which is rather near to the limiting value
$ \alpha = 1/2 $.
In other words, large values of $\alpha$ are expected in typical
inflationary scenarios driven by bulk viscosity.

Closely related to the bulk viscous pressure is the phenomenon of
particle production since the latter can be phenomelogically
interpreted as bulk viscosity \cite{Hu}, \cite{Zeldovich} and therefore
our above result involving $\pi$ finds a natural extension to
inflationary models driven either by cosmological particle production
(due to the interaction between the gravitational field and the quatum
vacuum) or by the decay of massive particles of the primordial
plasma into lighter ones. In particular there is a well-known
relationship between this dissipative pressure and the rate of particle
production $\Gamma$ for isotropic cosmological expansions when the
former is assumed to be adiabatic \cite{zimdahl}
\begin{equation}
\pi = - (\rho + p) \frac{\Gamma}{3H}.
\label{gammapi}
\end{equation}

In this connection Pav\'{o}n {\it et al.} \cite{PaGaLD} have considered the
decay of a non-relativistic fluid of massive particles with energy density
$\rho_{1}$ into a radiation-like fluid, with energy density
$\rho_{2}$ proportional to the fourth power of temperature.
Then assuming the relaxation time to be of the order of magnitude of the
mean free interaction time between
the matter and radiation particles, it follows that
\begin{equation}
\alpha=\frac{\beta}{2\left(\frac{4}{3}+\frac{\rho_1}{\rho_2}\right)}
\label{alb}
\end{equation}
\noindent
(note that our $\beta$ corresponds to their $\alpha$).

From (\ref{alb}) we have that
$\alpha$ grows with $\beta$, i.e. with the bulk viscosity pressure,
thereby larger values of $\alpha$ are related to stronger expansions.
This is only natural since for expanding universes the latter pressure
implies a negative contribution to the total fluid pressure and therefore
helps to accelerate the expansion.

Our next exemple includes two family of models proposed by Barrow
\cite{Ba} based in an effective bulk viscous pressure related to the
very high rate of quantum production of fundamental strings \cite{Tu}.
These two families correspond to string--driven inflationary and
deflationary models depending on if the Universe evolves from
a non--inflationary stage to an inflationary one or from a de Sitter
phase to a Friedmann expansion, respectively.  Barrow assumes
a spatially--flat scenario governed by
\[
3 \, H^{2} = \rho ,
\]
\noindent
and with the viscous pressure given by $\pi = -3 \, \zeta H $.
For $\zeta\propto\rho^{m}$ ($m=$constant), Barrow obtains
when $m<1/2$ and $H \geq H_{0} =$ constant, isotropic cosmological
models which exhibit inflationary behaviour. They begin at a
Friedmann singularity and approach from above a de Sitter state
as $t \rightarrow \infty$.

If $m>1/2$ and $H \leq H_{0}$ deflationary  expansions follow. These
begin in a de Sitter state with $H=H_{0}$ and evolve towards a Friedmann
asymptote. In both kind of models one has $\dot{H}<0$.

Now, assuming as in \cite{PaGaLD} $\tau_{\zeta}\propto{H^{-1}}$ and for
$\zeta$ the power-law dependence upon the density mentioned above, it is
easily obtained from (\ref{alfag}) with $\kappa = \eta = 0$
\begin{equation}
\alpha\propto H^{2m-1},
\label{alor}
\end{equation}
where (\ref{g}) has been used.

Thus, for $m>1/2$, $\alpha$ decreases during the expansion, explaining
thereby the deflationary behaviour of the models.
For $m<1/2$, $\alpha$ increases as $H$ decreases, leading to stronger
expansion (inflation).
This suggests that any inflationary model (driven by dissipation) endowed
with a mechanism for achieving an exit from inflation (a desirable feature),
should allow for changing from an increasing to a decreasing $\alpha$,
during the evolution. All these results confirm the role of $\alpha$ as
a measure of the ``strength'' of the expansion mentioned above.

It is worth stressing again that in the more general case (heat
flux plus shear and bulk viscosity), $\alpha$ may be larger than unity
without violating causality conditions \cite{HeMa98}.
This suggests that one may build more ``efficient'' inflationary models
by resorting to a combination of all forms of dissipation
(heat current plus either viscosity or particle production or both).
Obviously, the difficulty there would be
to deal with a more general geometry since to accommodate an energy-flux
together with shear stress the spacetime must be both inhomogeneous and
anisotropic.

Before closing we would like to emphasize that bulk viscosity may also
be interpreted as the effect of some scalar field, say $\phi$, which
has found a more ample audience as a candidate to drive the very
early accelerated expansion invoked  by inflationary models. The
connection between these two disparate quantities,
$\pi = - \Gamma \, \dot{\phi}^{2}/(3 H)$,
follows from specializing equation (\ref{gammapi}) to the scalar
field case only that in this instance $\Gamma$ is the decay rate of
$\phi$. This has been used to phenomelogically model the reheating
phase of the universe right after inflation \cite{zimdahl94}.
Although the behaviour of the corresponding control parameter
$\alpha$ may also be studied following parallel lines to that
of above, the involved task is by no means trivial and we
leave it to a future research.

\section*{Acknowledgements} This work has been partially supported by
the Spanish Ministry of Eductaion under grant PB94-0718.

\end{document}